\documentclass[12pt]{article}

\usepackage{graphics}

\title{A Mean Atom Trajectory Model for Monatomic Liquids}
\author{Eric D.\ Chisolm and Duane C.\ Wallace \\ Theoretical Division \\
        Los Alamos National Laboratory \\ Los Alamos, NM~~87545}

\begin{document}

\maketitle

\begin{abstract}
A recent description of the motion of atoms in a classical monatomic
system in liquid and supercooled liquid states divides the motion into
two parts: oscillations within a given many-particle potential valley,
and transit motion which carries the system from one many-particle
valley to another.  Building on this picture, we construct a model for
the trajectory of an average atom in the system.  The trajectory
consists of oscillations at the normal-mode distribution of
frequencies, representing motion within a fluctuating single-particle
well, interspersed with position- and velocity-conserving transits to
similar adjacent wells.  For the supercooled liquid in nondiffusing
states, the model gives velocity and displacement autocorrelation
functions which exactly match those found in the many-particle
harmonic approximation, and which are known to agree almost precisely
with molecular dynamics (MD) simulations of liquid Na.  At higher
temperatures, by allowing transits to proceed at a
temperature-dependent rate, the model gives velocity autocorrelation
functions which are also in remarkably good agreement with MD
simulations of Na at up to three times its melting temperature.  Two
independent processes in the model relax velocity autocorrelations:
(a) dephasing due to the presence of many frequency components, which
operates at all temperatures but which produces zero diffusion, and
(b) the transit process, which increases with increasing temperature
and which produces diffusion.  Compared to several treatments of
velocity autocorrelations based on instantaneous normal modes, the
present model offers an advantage:  It provides a single-atom
trajectory in real space and time, including transits and valid for
arbitrary times, from which all single-atom correlation functions can
be calculated, and they are also valid at all times.
\end{abstract}

\section{Introduction}
\label{intro}

The motion of atoms in a liquid can be divided into two constituent
\mbox{parts}: (a) oscillation in a valley of the liquid's many-body
potential and (b) transits between many-body valleys.  The latter
process is responsible for self-diffusion.  As we showed in a previous
paper \cite{chis}, which drew upon earlier results of Clements and
Wallace \cite{enkin, radang}, the former motion can be modeled very
precisely under the assumption that the valleys are nearly harmonic,
with the majority of valleys (the random valleys) sharing a common
spectrum of frequencies.  Specifically, a purely harmonic model
provides an extremely accurate formula for $\hat{Z}(t)$, the
normalized velocity autocorrelation function, in the nondiffusing
regime.  Here we will use this picture and our previous work to
justify a ``mean atom trajectory'' model, a single-atom model that
approximates the behavior of an average atom in the liquid and
correctly reproduces its nondiffusing behavior.  Then we will
introduce a simple intuitive account of the transit process that
allows us to extend the model to the self-diffusing regime.  In Sec.\
\ref{model} we develop this model and explain how it is used to
calculate $\hat{Z}(t)$ for a diffusing liquid.  Then we fit the model
to MD simulations at various temperatures in Sec.\ \ref{MD}, and we
comment on the quality of the results.  In Sec.\ \ref{concl} we
compare the present mean atom trajectory model with previous work
based on Instantaneous Normal Modes (INM), and with an earlier
independent atom model \cite{oldvacf}, and we summarize our
conclusions.

\section{The Mean Atom Trajectory Model}
\label{model}

\subsection{General comments}
\label{general}

To form an appropriate basis for our model, we must begin with some
initial reasonable approximations about the nature of the valleys and
transits in the real liquid.  As mentioned in the Introduction,
available evidence suggests that the valleys are nearly harmonic, and
we will continue to assume that here.  Further, we will assume that
transits between valleys occur instantaneously and are local in
character; that is, each transit involves only a few neighboring
atoms.  

Now for the nondiffusing supercooled liquid, there are no transits,
and as we've seen the system's motion is accurately expressed in terms
of the harmonic normal modes.  But it is also legitimate to consider
this motion from the point of view of a single atom, and when we do so
we see that each atom moves along a complicated trajectory within a
single-particle potential well which fluctuates because of the motion of 
neighboring atoms, but whose center is fixed in space.  An
important timescale for this motion is the single-atom mean
vibrational period $\tau$, which we take as $2 \pi / {\omega_{\rm
rms}}$, where ${\omega_{\rm rms}}$ is the rms frequency of the set of
normal modes.  Let us now follow the single-atom description as the
temperature is increased.  Once the glass transition is passed, the
atom will begin to make transits from one single-particle well to
another \cite{glass}, and even before the melting temperature is
reached, the transit rate will be on the order of one per mean
vibrational period.  Since each atom has approximately ten neighbors,
roughly ten transits will occur in its immediate vicinity every
period, changing the set of normal mode eigenvectors each time; thus
we conclude that a decomposition of the motion into normal modes will
not be useful when the liquid is diffusing, so we have no choice but
to follow a single-atom description for the liquid state
\cite{oldvacf}.  That being the case, we shall start from the
beginning with a single-atom description, in order to construct a
unified model for diffusing and nondiffusing motion alike.

\subsection{Nondiffusing regime}
\label{nondiff}

Our starting point will be in the nondiffusing regime, where a normal
mode analysis is still valid.  As shown in \cite{chis}, the $i$th
coordinate of the $K$th atom in an $N$-body harmonic valley can be
written
\begin{equation}
u_{Ki}(t) = \sum_{\lambda} w_{Ki, \lambda} a_{\lambda}
            \sin(\omega_{\lambda}t + \alpha_{\lambda}),
\label{uharmoft}
\end{equation}
where the $w_{Ki,\lambda}$ form a $3N \times 3N$ orthogonal matrix,
the $\omega_{\lambda}$ are the frequencies of the normal modes, and
the $a_{\lambda}$ are the amplitudes of the modes.  Three of the
modes have zero frequency and correspond to center of mass motion; here
we demand that the center of mass is stationary, so the system has
only $3N-3$ independent degrees of freedom, the zero frequency modes
are absent, and the sum over $\lambda$ runs from 1 to $3N-3$.  To
make this an equation for a ``mean'' atom, we first drop the index
$K$:
\begin{equation}
u_{i}(t) = \sum_{\lambda} w_{i \lambda} a_{\lambda}
            \sin(\omega_{\lambda}t + \alpha_{\lambda}).
\label{u1oft}
\end{equation}
Now we must reinterpret the $w_{i \lambda}$ since they no longer form an 
orthogonal (or even square) matrix.  Let
\begin{equation}
\mbox{\boldmath $w$}_{\lambda} = w_{1 \lambda} \hat{\mbox{\boldmath $x$}} + 
                                 w_{2 \lambda} \hat{\mbox{\boldmath $y$}} + 
                                 w_{3 \lambda} \hat{\mbox{\boldmath $z$}}
\label{defw}
\end{equation}
so
\begin{equation}
\mbox{\boldmath $u$}(t) = \sum_{\lambda} \mbox{\boldmath $w$}_{\lambda} 
                          a_{\lambda} \sin(\omega_{\lambda}t + 
                          \alpha_{\lambda}).
\label{u2oft}
\end{equation}
We will ultimately consider situations in which the well center is
allowed to move, so let \mbox{\boldmath $r$}$(t)$ be the atom's
position, \mbox{\boldmath $R$} be the location of the center of
the well, and \mbox{\boldmath $u$}$(t)$ be the atom's displacement
from the well center; then
\begin{equation}
\mbox{\boldmath $r$}(t) = \mbox{\boldmath $R$} + \sum_{\lambda} 
                          \mbox{\boldmath $w$}_{\lambda} a_{\lambda} 
                          \sin(\omega_{\lambda}t + \alpha_{\lambda})
\label{u3oft}
\end{equation}
with velocity
\begin{equation}
\mbox{\boldmath $v$}(t) = \sum_{\lambda} \mbox{\boldmath
  $w$}_{\lambda} a_{\lambda} \omega_{\lambda} \cos(\omega_{\lambda}t +
  \alpha_{\lambda}).
\label{v3oft}
\end{equation}

We now have the basic formula, but we must decide how to assign values
to the {\boldmath $w$}$_{\lambda}$, $a_{\lambda}$, and
$\alpha_{\lambda}$.  Let's do this by calculating $Z(t)$, the velocity
autocorrelation function, in this model and comparing it to the
harmonic result derived in \cite{chis}.
\begin{eqnarray}
Z(t) & = & \frac{1}{3} \langle \mbox{\boldmath $v$}(t) \cdot
           \mbox{\boldmath $v$}(0) \rangle \nonumber \\ 
& = & \frac{1}{3} \sum_{\lambda \lambda'} \mbox{\boldmath $w$}_{\lambda} 
      \cdot \mbox{\boldmath $w$}_{\lambda'} a_{\lambda} a_{\lambda'}
      \omega_{\lambda} \omega_{\lambda'} \langle \cos(\omega_{\lambda}t +
      \alpha_{\lambda}) \cos(\alpha_{\lambda'}) \rangle
\label{1stZ}
\end{eqnarray}
Let's assign the $\alpha_{\lambda}$ randomly and average over each 
$\alpha_{\lambda}$ {\em separately}; then if $\lambda \neq \lambda'$,
\begin{equation}
\langle \cos(\omega_{\lambda}t + \alpha_{\lambda}) \cos(\alpha_{\lambda'}) 
\rangle = \langle \cos(\omega_{\lambda}t + \alpha_{\lambda})  \rangle 
\langle \cos(\alpha_{\lambda'}) \rangle = 0 
\end{equation}
but if $\lambda = \lambda'$, then
\begin{equation}
\langle \cos(\omega_{\lambda}t + \alpha_{\lambda}) \cos(\alpha_{\lambda'}) 
\rangle = \langle \cos(\omega_{\lambda}t + \alpha_{\lambda}) 
\cos(\alpha_{\lambda}) \rangle = \frac{1}{2} \cos(\omega_{\lambda}t).
\end{equation}
Thus the $\lambda \neq \lambda'$ terms in Eq.\ (\ref{1stZ}) are eliminated and 
we have
\begin{equation}
Z(t) = \frac{1}{6} \sum_{\lambda} | \mbox{\boldmath $w$}_{\lambda} |^2 
       a_{\lambda}^2 \omega_{\lambda}^2 \cos(\omega_{\lambda}t).
\label{2ndZ}
\end{equation}
Note that the formula for $Z(t)$ in the harmonic theory also lacks 
off-diagonal terms \cite{chis}, but for a different reason:  The 
orthogonality of the matrix $w_{Ki,\lambda}$ removes them in that case. 
Now let us make the assignment
\begin{equation}
\mbox{\boldmath $w$}_{\lambda} = \frac{1}{\sqrt{N-1}} \hat{\mbox{\boldmath 
                                 $w$}}_{\lambda}
\label{defwhat}
\end{equation}
where $\hat{\mbox{\boldmath $w$}}_{\lambda}$ is a randomly chosen unit vector; 
then
\begin{equation}
Z(t) = \frac{1}{6N-6} \sum_{\lambda} a_{\lambda}^2 \omega_{\lambda}^2 
       \cos(\omega_{\lambda}t).
\label{3rdZ}
\end{equation}
This is the same expression for $Z(t)$ that one derives in the
harmonic model (see Eq.\ (10) of \cite{chis}) for {\em any}
distribution of normal mode amplitudes $a_{\lambda}$, not just the
thermal equilibrium distribution.  Thus our model with these choices
of $\alpha_{\lambda}$ and \mbox{\boldmath $w$}$_{\lambda}$ correctly
reproduces $Z(t)$ for any equilibrium or nonequilibrium ensemble in
the harmonic theory.  To recover the equilibrium result, we make the
final substitution
\begin{equation}
a_{\lambda} = \sqrt{\frac{2kT}{M \omega_{\lambda}^2}}
\end{equation}
with the result
\begin{equation}
Z(t) =  \frac{1}{3N-3} \frac{kT}{M} \sum_{\lambda} \cos(\omega_{\lambda}t),
\label{4thZ}
\end{equation}
which is Eq.\ (13) from \cite{chis}.  Thus our model assumes that the 
motion of a mean nondiffusing atom in thermal equilibrium at temperature $T$ 
is given by
\begin{equation}
\mbox{\boldmath $r$}(t) = \mbox{\boldmath $R$} + \frac{1}{\sqrt{N-1}}
                          \sqrt{\frac{2kT}{M}} \sum_{\lambda} 
                          \hat{\mbox{\boldmath $w$}}_{\lambda}
                          \omega_{\lambda}^{-1} \sin(\omega_{\lambda}t + 
                          \alpha_{\lambda})
\label{u4oft}
\end{equation}
where the phases $\alpha_{\lambda}$ and unit vectors
$\hat{\mbox{\boldmath $w$}}_{\lambda}$ are randomly chosen.  

By construction this model gets the same result for $Z(t)$, and thus
$\langle v^2 \rangle$, as the harmonic model; does it correctly
reproduce any other functions?  We can check by calculating the analog
of $Z(t)$ for positions, $\langle \mbox{\boldmath $u$}(t) \cdot
\mbox{\boldmath $u$}(0) \rangle$.  Using Eq.\ (\ref{uharmoft}) for
$u_{Ki}(t)$ and calculating for the harmonic model as in \cite{chis},
and using Eq.\ (\ref{u1oft}) for \mbox{\boldmath $u$}$(t)$ and
calculating as above, one finds in both cases that
\begin{equation}
\langle \mbox{\boldmath $u$}(t) \cdot \mbox{\boldmath $u$}(0) \rangle = 
\frac{1}{2N-2} \sum_{\lambda} a_{\lambda}^2 \cos(\omega_{\lambda}t).
\end{equation}
Again the substitution $a_{\lambda} = \sqrt{2kT/M\omega_{\lambda}^2}$ 
recovers the correct thermal equilibrium result.  So this model correctly 
reproduces $\langle \mbox{\boldmath $u$}(t) \cdot \mbox{\boldmath $u$}(0) 
\rangle$ and $\langle u^2 \rangle$ as well.  Notice that because we have a 
closed form for \mbox{\boldmath $u$}$(t)$, all of the correlation functions 
calculated so far also have a closed form.  Once we introduce transits, this 
will no longer be the case.

\begin{figure}
\includegraphics{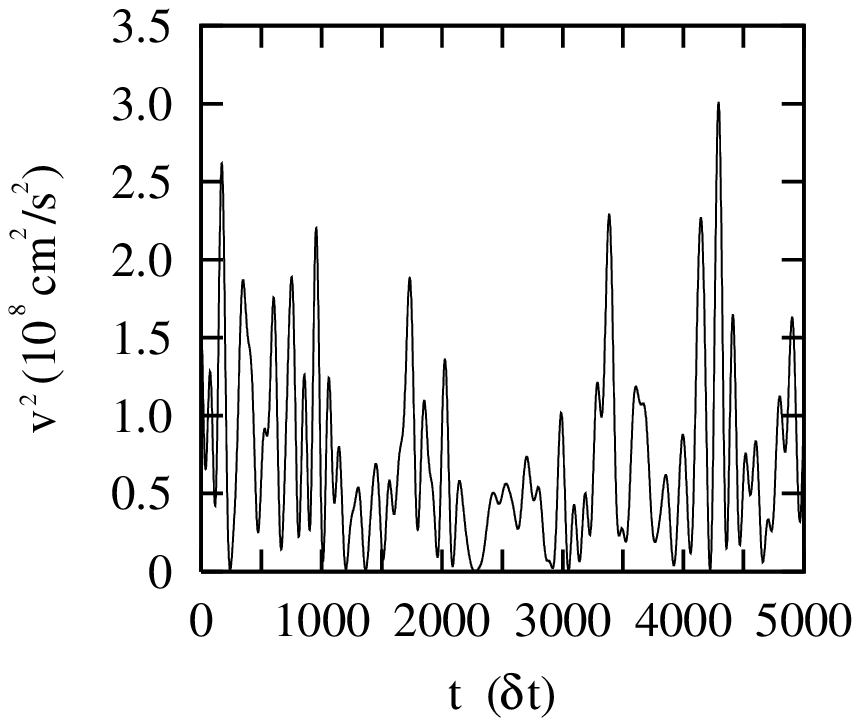}
\includegraphics{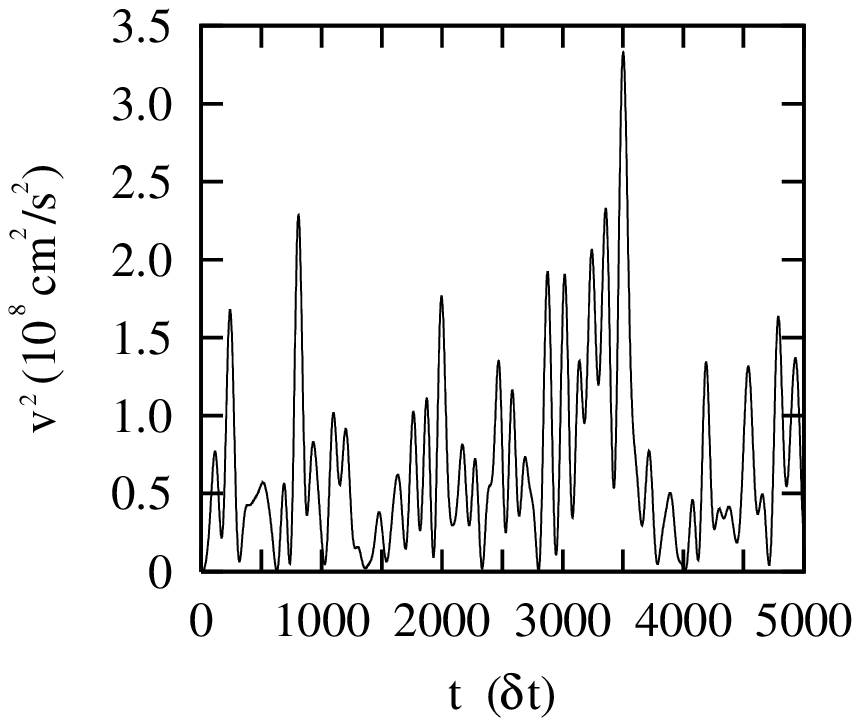}
\caption{$v^2$ for a randomly chosen atom in an MD simulation of Na at 6.69 K 
         (top) is compared with $v^2$ for the mean atom trajectory model in the
         nondiffusing regime (bottom) over 5000 timesteps of duration $\delta 
         t = 1.4 \times 10^{-15}$ s.  Notice that both fluctuations have 
         roughly the same magnitude and oscillate at approximately the same 
         frequencies.}
\label{vsqvsMD}
\end{figure}

While this model compares well with the harmonic model, one might wonder how 
well it compares with molecular dynamics (MD) results.  In Fig.\ \ref{vsqvsMD} 
a graph of $v^2$ as a function of $t$ for the model in equilibrium at 6.69 K is
compared to a randomly chosen particle from an MD run
of liquid Na also at 6.69 K, a temperature at which it is known the
sample is nondiffusing (see \cite{chis} for details).  Note that in
both graphs $v^2$ has approximately the same amplitude, and the gaps
between peaks are roughly the same size, indicating oscillations at
frequencies in the same ranges.  Thus not only functions of the motion
but the motion itself shows strong qualitative agreement with MD in
the nondiffusing regime.

The MD system with which we will be comparing our model has $N=500$
particles, so it has 1497 normal mode frequencies.  (Remember that the
zero frequency modes are removed at the start.)  When we introduce
transits, we will have to evaluate $Z(t)$ numerically, and for speed
of computation we would like to use only a representative subset of the normal 
mode frequencies, say 75 instead of all 1497.  The original set of
frequencies is determined as described in \cite{chis}.  We decided
which subset to use by calculating three moments of the full frequency
distribution defined below:
\begin{eqnarray}
\omega_{2} & = & \left[ \frac{5}{3} \langle \omega_{\lambda}^2 \rangle
                 \right]^{1/2} \nonumber \\
\omega_{0} & = & \exp \langle \ln(\omega_{\lambda}) \rangle \nonumber \\
\omega_{-2} & = & \left[ \frac{1}{3} \langle \omega_{\lambda}^{-2} \rangle
                   \right]^{-1/2}.
\end{eqnarray}
Note that by this definition $\omega_{\rm rms} = \sqrt{3/5}\, \omega_{2}$. 
We then calculated the same three moments for several sets of 75 frequencies 
evenly spaced throughout the full set, and we chose the set that best fit the 
moments to use in further calculations.  This process is somewhat subjective, 
because skewing the sample in favor of lower frequencies improves the accuracy 
of $\omega_{-2}$ but reduces the accuracy of $\omega_{2}$, and the opposite is 
true if one skews in favor of high frequencies.  The table below shows the 
values of the moments for the full set and the reduced set of 75 that we 
ultimately chose.  The frequencies are in units of $\delta t^{-1}$, where 
$\delta t = 1.4 \times 10^{-15}$ s is the timestep of our MD simulations 
\cite{chis}.
\begin{center}
\begin{tabular}{ccc}
              &  Full   & Reduced \\ \hline
$\omega_{2}$  & 0.02826 & 0.02824 \\
$\omega_{0}$  & 0.01807 & 0.01803 \\
$\omega_{-2}$ & 0.02101 & 0.02111
\end{tabular}
\end{center}
To use the reduced set we must rewrite our formulas for 
\mbox{\boldmath $r$}$(t)$ and \mbox{\boldmath $v$}$(t)$ \mbox{slightly}.  Let 
$\Lambda$ be the total number of frequencies; then $\Lambda = 3N - 3$ for the 
full set so \mbox{\boldmath $r$}$(t)$ and \mbox{\boldmath $v$}$(t)$ become
\begin{eqnarray}
\mbox{\boldmath $r$}(t) & = & \mbox{\boldmath $R$} + \sqrt{\frac{3}{\Lambda}}
                              \sqrt{\frac{2kT}{M}} \sum_{\lambda=1}^{\Lambda} 
                              \hat{\mbox{\boldmath $w$}}_{\lambda} \,
                              \omega_{\lambda}^{-1} \sin(\omega_{\lambda}t + 
                              \alpha_{\lambda}) \nonumber \\
\mbox{\boldmath $v$}(t) & = & \sqrt{\frac{3}{\Lambda}} \sqrt{\frac{2kT}{M}} 
                              \sum_{\lambda=1}^{\Lambda} 
                              \hat{\mbox{\boldmath $w$}}_{\lambda}
                              \cos(\omega_{\lambda}t + \alpha_{\lambda}).
\label{uoft}
\end{eqnarray}
This form is also correct for the reduced set of frequencies
$\omega_{\lambda}$, so this is the form we will use.  As another check
on the accuracy of our results with only 75 frequencies, we
recalculated the original $\sum \cos(\omega_{\lambda}t)$ expression
for $\hat{Z}(t)$ using both the full and reduced sets; their
disagreement is at most 0.01 out to time $1000 \, \delta t$.  However,
the discrepancy grows beyond that time, as the expression using the
reduced set begins to experience revivals.  Hence we will not consider
$\hat{Z}(t)$ beyond that point.

\subsection{Diffusing regime}
\label{diff}

To incorporate diffusion into the mean atom trajectory model, we rely on 
Wallace's notion of a single-particle transit \cite{liqdyn}, a nearly
instantaneous transition from one well to another.  As discussed in 
\cite{glass}, we expect transits to be governed not by thermal activation 
(having enough energy to escape a fixed well) but by {\em correlations} 
(neighbors must be positioned properly for a low-potential path to open between
wells).  We implement this property by having transits occur at a temperature 
dependent {\em rate} $\nu(T)$, so in a small time interval $\Delta t$ the 
probability of a single transit is $\nu \Delta t$.

We model the transit process itself by assuming it occurs
instantaneously in the forward direction; from this we can determine
the parameters {\boldmath $R$}, $\hat{\mbox{\boldmath
$w$}}_{\lambda}$, and $\alpha_{\lambda}$ appearing in Eq.\
(\ref{uoft}) after the transit in terms of the same quantities before
the transit.  Since the process is instantaneous, both {\boldmath
$r$}$(t)$ and {\boldmath $v$}$(t)$ are the same before and afterwards.
Let {\boldmath $R$}$^{\rm before}$ and {\boldmath $R$}$^{\rm after}$
be the well centers before and after the transit and let {\boldmath
$u$}$^{\rm before}$ and {\boldmath $u$}$^{\rm after}$ be the
corresponding displacements from the well centers.  Then
$\mbox{\boldmath $r$}^{\rm before} = \mbox{\boldmath $r$}^{\rm after}$
implies
\begin{equation}
\mbox{\boldmath $R$}^{\rm before} + \mbox{\boldmath $u$}^{\rm before} = 
\mbox{\boldmath $R$}^{\rm after} + \mbox{\boldmath $u$}^{\rm after}.  
\end{equation}
To transit {\em forward}, we assume the center of the new well lies along the 
line between the old well center and the atom, but it lies on the opposite side
of the atom from the old well center an equal distance away.  This implies
{\boldmath $u$}$^{\rm after} = -\mbox{\boldmath $u$}^{\rm before}$, so 
\begin{equation}
\mbox{\boldmath $R$}^{\rm before} + \mbox{\boldmath $u$}^{\rm before} = 
        \mbox{\boldmath $R$}^{\rm after} - \mbox{\boldmath $u$}^{\rm before} 
\end{equation}
with the result
\begin{equation}
\mbox{\boldmath $R$}^{\rm after} =  \mbox{\boldmath $R$}^{\rm before} + 
                                    2\mbox{\boldmath $u$}^{\rm before}. 
\label{Rafttrans}
\end{equation}
This determines the new well center in terms of the coordinates before the 
transit.  As for the unit vectors $\hat{\mbox{\boldmath $w$}}_{\lambda}$, since
they are randomly generated and play no role in calculating $\hat{Z}(t)$, we 
have decided to leave them unchanged by transits.  Finally, we have the phases 
$\alpha_{\lambda}$.  We must use these to implement the relations
\begin{equation}
\mbox{\boldmath $u$}^{\rm after} = -\mbox{\boldmath $u$}^{\rm before}, \ \ 
\mbox{\boldmath $v$}^{\rm after} = \mbox{\boldmath $v$}^{\rm before}
\end{equation}
which we have assumed above.  Since {\boldmath $u$}$(t)$ is a sum of sines and 
{\boldmath $v$}$(t)$ a sum of cosines, the simplest way to change the sign of 
{\boldmath $u$}$(t)$ while preserving that of {\boldmath $v$}$(t)$ is to 
reverse the signs of the arguments $(\omega_{\lambda}t + \alpha_{\lambda})$
in Eq.\ (\ref{uoft}).  Let the transit occur at time $t_{0}$; then
\begin{equation}
\omega_{\lambda}t_{0} + \alpha_{\lambda}^{\rm after} = -(\omega_{\lambda}t_{0} 
                      + \alpha_{\lambda}^{\rm before})
\end{equation}
so
\begin{equation}
\alpha_{\lambda}^{\rm after} = -2\omega_{\lambda}t_{0} - \alpha_{\lambda}^{\rm 
                                before}.
\end{equation}
Thus, in this model a transit is implemented at time $t_{0}$ by leaving the 
$\hat{\mbox{\boldmath $w$}}_{\lambda}$ alone and making the substitutions
\begin{eqnarray}
\mbox{\boldmath $R$} & \rightarrow & \mbox{\boldmath $R$} + 2\mbox{\boldmath 
                                                 $u$}(t_{0}) \nonumber \\
\alpha_{\lambda} & \rightarrow & -2\omega_{\lambda}t_{0} - \alpha_{\lambda}.
\label{imptrans}
\end{eqnarray}
This conserves {\boldmath $r$}, reverses the sign of {\boldmath $u$}, and 
conserves {\boldmath $v$}. 

Now our mean atom trajectory model consists of nondiffusive motion
between transits as given by Eq.\ (\ref{uoft}), with a given
probability in each small time interval that {\boldmath $R$} and the
phases $\alpha_{\lambda}$ will be replaced with new values as
determined in Eq.\ (\ref{imptrans}).  The addition of transits means
that we no longer have closed form expressions for {\boldmath
$r$}$(t)$ and {\boldmath $v$}$(t)$ for all times, so we have no closed
form expression for $\hat{Z}(t)$; but this model can be implemented
easily on a computer in a manner analogous to an MD calculation, and in
that way we can calculate autocorrelation functions.  We turn to that
calculation next.

\subsection{Evaluating $\hat{Z}(t)$ in the diffusing regime}
\label{evalZ}

To calculate $\hat{Z}(t)$, we select a value for the rate $\nu$,
generate a random set of $\hat{\mbox{\boldmath $w$}}_{\lambda}$ and
$\alpha_{\lambda}$, and use Eq.\ (\ref{uoft}) to calculate {\boldmath
$r$}$(t)$ and {\boldmath $v$}$(t)$ from $t = 0$ to $t=t_{\rm max}$ in
increments of $\delta t$, where the criterion for choosing $t_{\rm
max}$ is discussed below and $\delta t$ is the timestep used in our MD
simulations (defined in Subsection \ref{nondiff}).  At each timestep,
we check to see if a transit occurs, and if so we implement Eq.\
(\ref{imptrans}) and continue with the new {\boldmath $R$} and
$\alpha_{\lambda}$.  We then calculate $\hat{Z}(t)$ using the formula
\begin{equation}
Z(t) = \frac{1}{3(t_{\rm max} - t) + 3} \sum_{t' = 0}^{t_{\rm max} - t} 
       \mbox{\boldmath $v$}(t + t') \cdot \mbox{\boldmath $v$}(t')
\end{equation}
and normalizing.  This equation is a modified form of the expression
used to calculate $Z(t)$ in MD; notice that the average over $t'$ has
the same effect as averaging separately over each phase
$\alpha_{\lambda}$ that appears in the velocity vectors.  Just as in
MD, we want to average over a large data set, so we require $t_{\rm
max} >> t$; we have chosen $t_{\rm max} = 20$ million timesteps and we
calculate $\hat{Z}(t)$ only to $t=1000$ timesteps.  We estimate the
total error from using only a subset of all 1497 frequencies and the
finite size of the data set to be at most 0.01; in particular, when
$\nu = 0$ the calculation converges to the closed form result $\sum
\cos(\omega_{\lambda}t)$ to this accuracy.

\section{Comparison with MD}
\label{MD}

\begin{figure}
\includegraphics{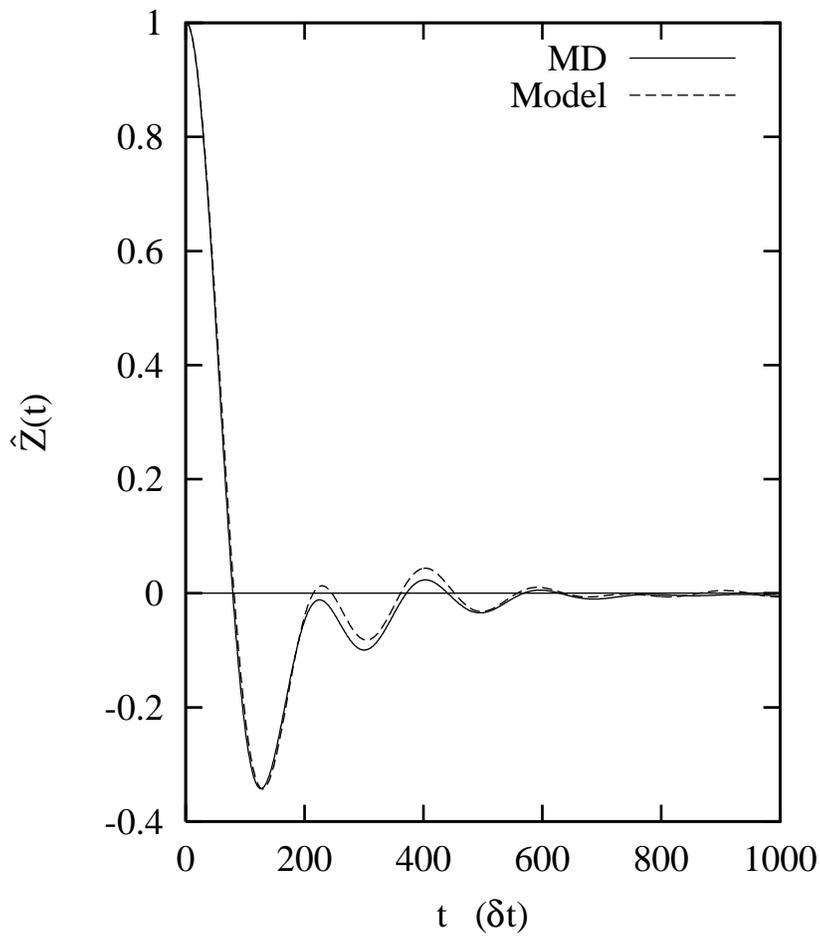}
\caption{The model prediction for $\hat{Z}(t)$ at $\nu = 0.35018 \, \tau^{-1}$ 
         compared with the MD result for supercooled liquid Na at $T = 
         216.3$ K.}
\label{216.3Kvsmodel}
\end{figure}

\begin{figure}
\includegraphics{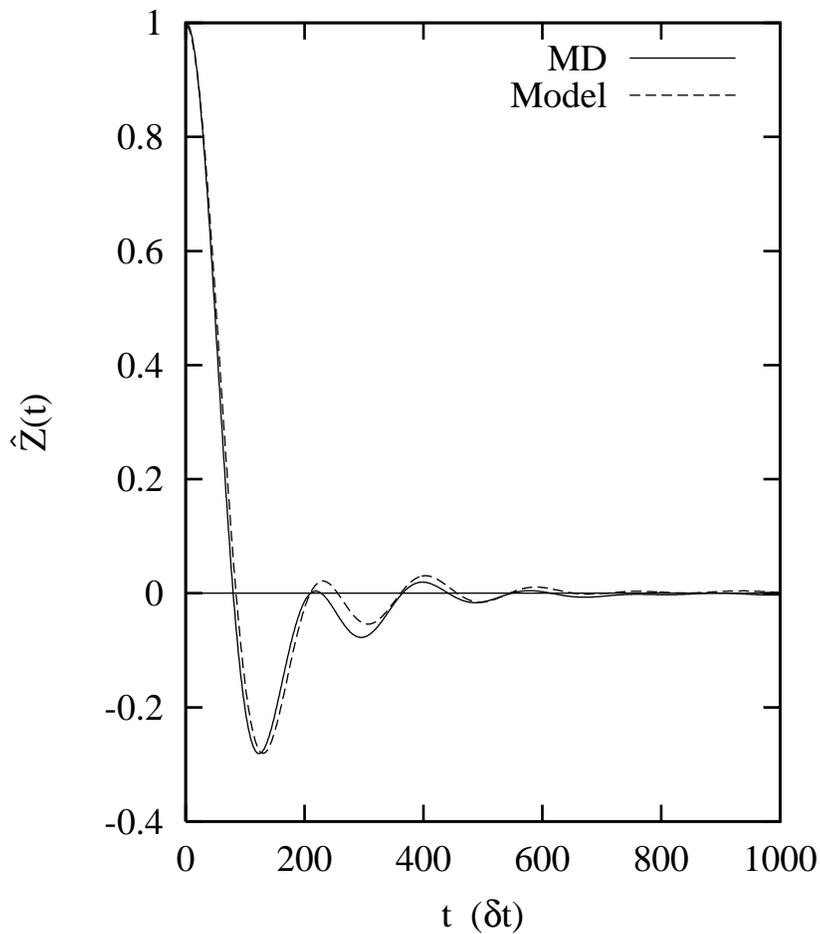}
\caption{The model prediction for $\hat{Z}(t)$ at $\nu = 0.60276 \, \tau^{-1}$ 
         compared with the MD result for supercooled liquid Na at $T = 
         309.7$ K.}
\label{309.7Kvsmodel}
\end{figure}

\begin{figure}
\includegraphics{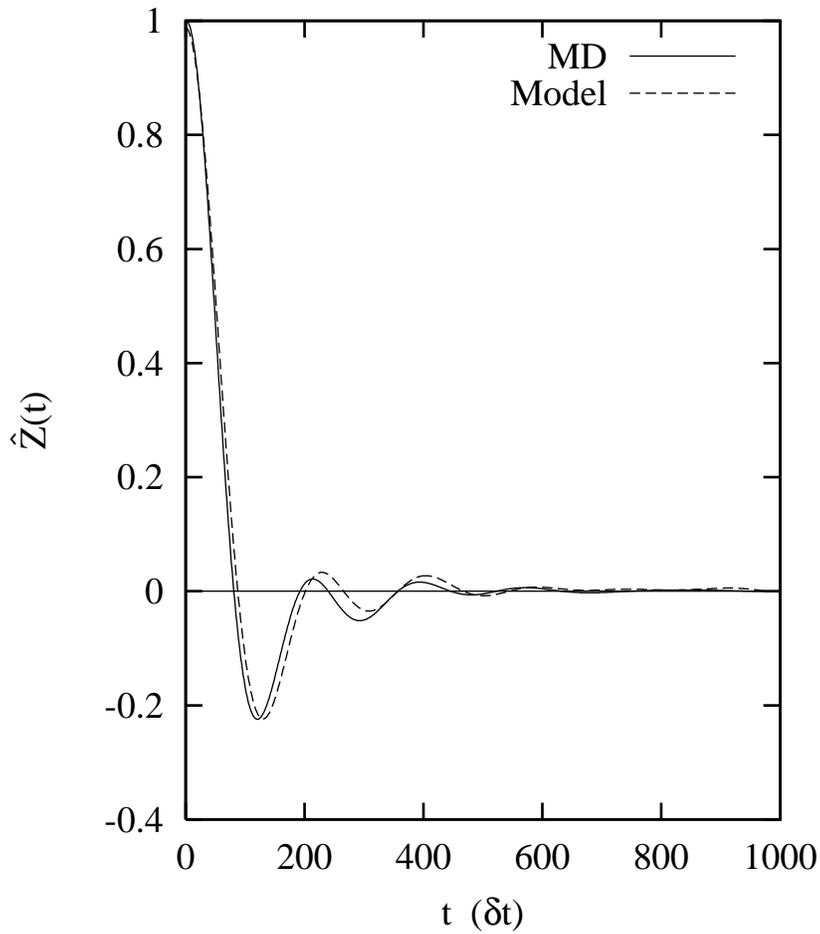}
\caption{The model prediction for $\hat{Z}(t)$ at $\nu = 0.83985 \, \tau^{-1}$ 
         compared with the MD result for liquid Na at $T = 425.0$ K.}
\label{425.0Kvsmodel}
\end{figure}

\begin{figure}
\includegraphics{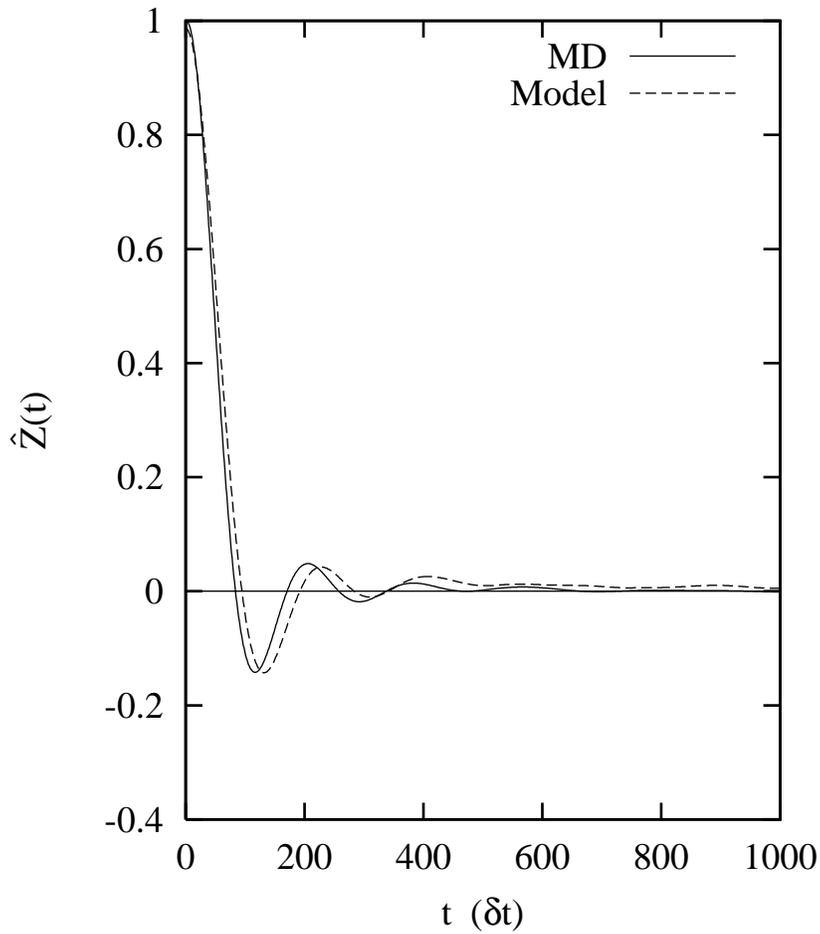}
\caption{The model prediction for $\hat{Z}(t)$ at $\nu = 1.24858 \, \tau^{-1}$ 
         compared with the MD result for liquid Na at $T = 664.7$ K.}
\label{664.7Kvsmodel}
\end{figure}

\begin{figure}
\includegraphics{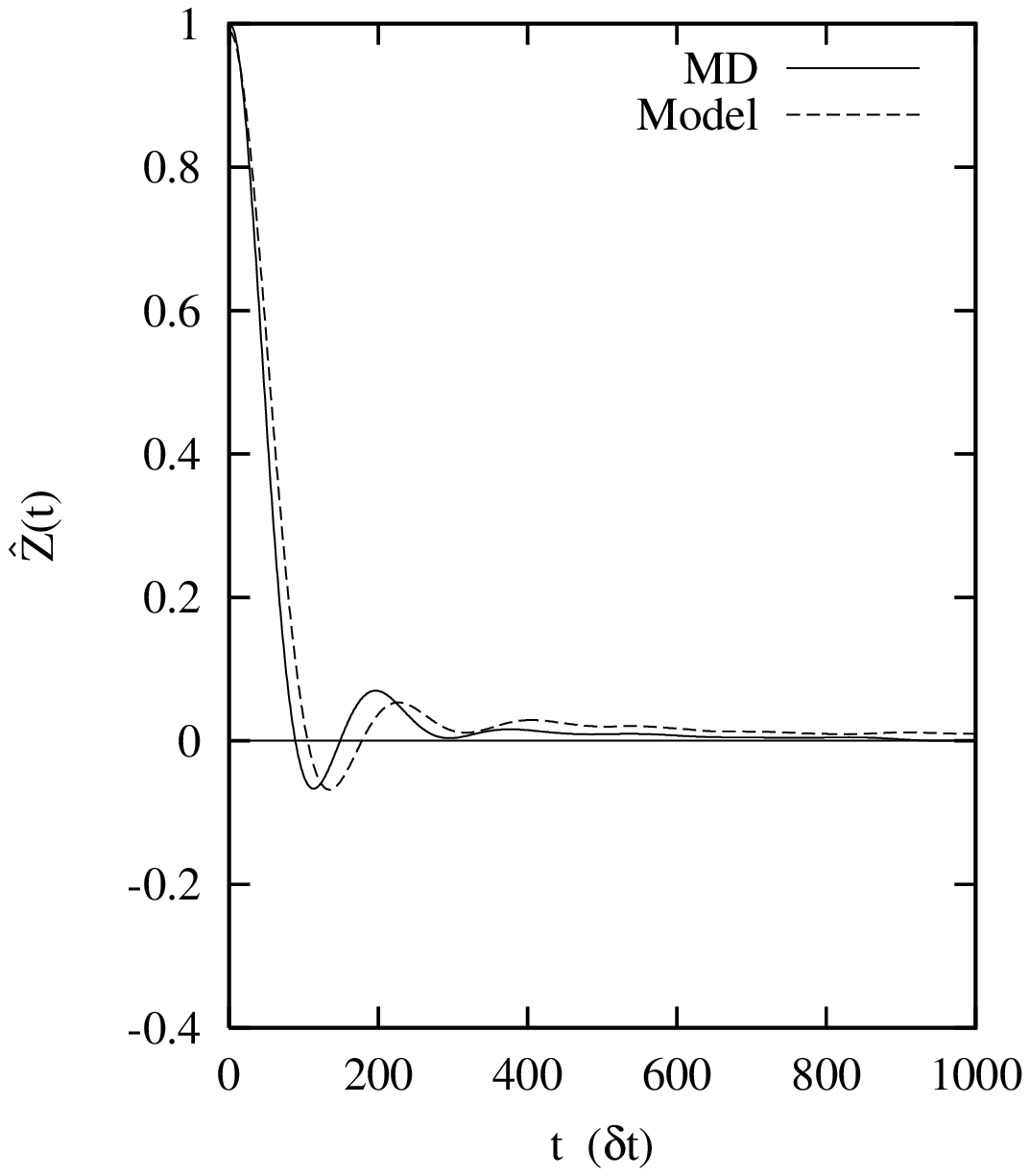}
\caption{The model prediction for $\hat{Z}(t)$ at $\nu = 1.68774 \, \tau^{-1}$ 
         compared with the MD result for liquid Na at $T = 1022.0$ K.}
\label{1022.0Kvsmodel}
\end{figure}

The MD setup with which we compared the predictions of this model is
the one described in \cite{chis}; $N=500$ atoms of Na move under the
influence of a highly realistic pair potential with the timestep
$\delta t = 1.4 \times 10^{-15}$ given in Subsection \ref{nondiff}.
We performed equilibrium runs of the system at 216.3 K, 309.7 K, 425.0
K, 664.7 K, and 1022.0 K, all temperatures at which the system is
diffusing.  Since $T_{m} = 371.0$ K for Na at this density, our
simulations range from the supercooled regime to nearly three times
the melting temperature.  We then ran the model for various values of
$\nu$, adjusting until the model matched the value of the first
minimum of $\hat{Z}(t)$ at each temperature.  The values of $\nu$ that
we fit for all temperatures are given below, and the resulting $\hat{Z}(t)$ 
for each $\nu$ is compared to the corresponding MD result in Figs.\ 
\ref{216.3Kvsmodel} through \ref{1022.0Kvsmodel}.  Here $\nu$ is 
expressed in units of $\tau^{-1}$ where $\tau = 2 \pi / \omega_{\rm 
rms}$ is the single-atom mean vibrational period defined in Subsection 
\ref{general}. 
\begin{center}
\begin{tabular}{rc}
$T$ (K) &  $\nu \ (\tau^{-1})$ \\ \hline
 216.3  & 0.35018 \\
 309.7  & 0.60276 \\
 425.0  & 0.83985 \\
 664.7  & 1.24858 \\
1022.0  & 1.68774
\end{tabular}
\end{center}
Notice that in all cases $\nu$ is of the same order of magnitude as
$\tau^{-1}$, indicating roughly one transit per mean vibrational
period, as mentioned in Subsection \ref{general}, and as predicted in
\cite{glass}.  

\begin{figure}
\includegraphics{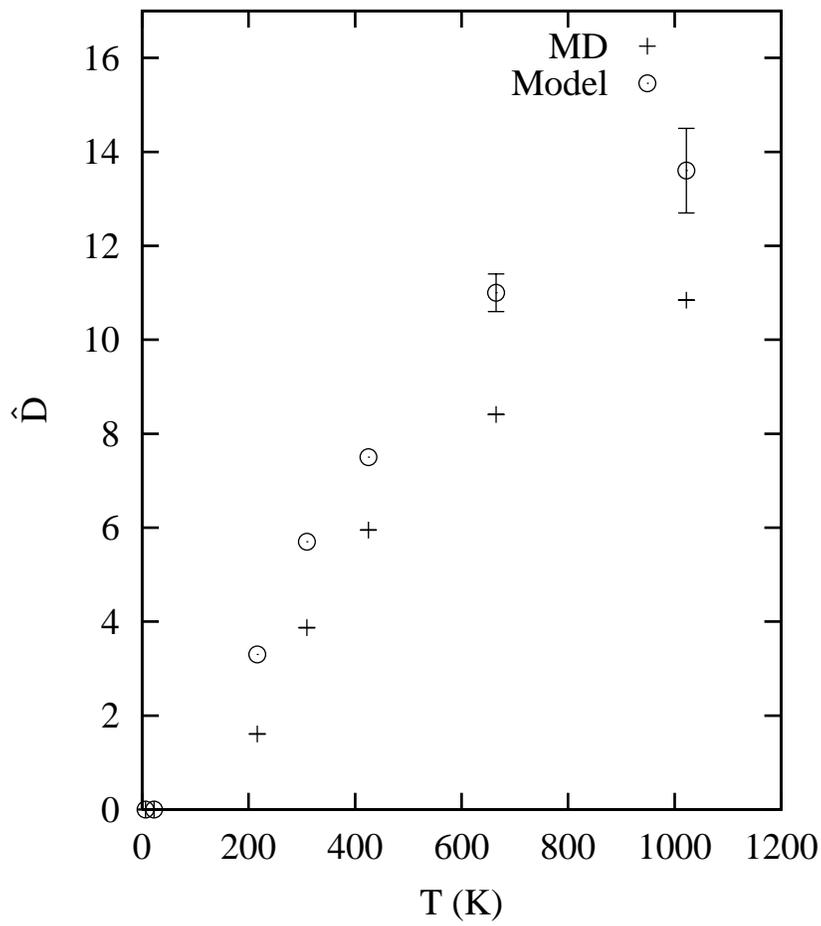}
\caption{$\hat{D}$ as a function of $T$ for both the model and MD.}
\label{DvsT}
\end{figure}

The most obvious trend exhibited by $\hat{Z}(t)$ from the five MD runs
is that its first minimum is rising with increasing $T$; as we mention
below, this is the primary reason for the increasing diffusion
coefficient $D$.  Note that the model is able to reproduce this most
important feature quite satisfactorily.  In fact, all five fits of
the model to the MD results capture their essential features, but we
do see systematic trends in the discrepancies.  First, note that the
location of the first minimum barely changes at all in the model as
$\nu$ is raised, but in MD the first minimum moves steadily to earlier
times as the temperature rises.  The first minimum occurs at a time
roughly equal to half of the mean vibrational period ($\tau = 287 \,
\delta t$ in this system), so the steady drift backward suggests that
the MD system is sampling a higher range of frequencies at higher $T$.
Also, for the three lowest temperatures the model tends to overshoot
the MD result in the vicinity of the first two maxima after the
origin, and at the highest two temperatures this overshoot is
accompanied by a positive tail that is slightly higher than the (still
somewhat long) tail predicted by MD.  These overshoots should clearly
affect the diffusion coefficient $D$, which is the integral of $Z(t)$.
To check this, we calculated the reduced diffusion coefficient
$\hat{D}$, the integral of $\hat{Z}(t)$, which is related to $D$ by
\begin{equation} D = \frac{kT}{M} \hat{D}. \end{equation}
The results are compared to the values of $\hat{D}$ calculated from
the MD runs in Fig.\ \ref{DvsT}.  The results from the two
nondiffusing runs discussed in \cite{chis} are also included.
In all of the diffusing cases, the model overestimates $\hat{D}$ by roughly 
the same amount, which we take to be the effect of the overshoots at the first 
two maxima.  At the higher temperatures the discrepancy is also higher, 
presumably due to the model's long tail.

It is interesting to note that this MD system produces results that
agree very closely with experiment: For example, the MD predicts that
at $T = 425.0$ K and $\rho = 0.925$ g/cm$^3$, $D = 6.40 \times
10^{-5}$ cm$^2$/s, while experiments by Larsson, Roxbergh, and Lodding
\cite{LRL} find that at $T = 425.0$ K and $\rho = 0.915$ g/cm$^3$, $D
= 6.020 \times 10^{-5}$ cm$^2$/s.  Hence our agreement with MD results
genuinely reflects agreement with properties of real liquid Na.

\section{Conclusions}
\label{concl}

We have presented a single-atom model of a monatomic liquid that
provides a unified account of diffusing and nondiffusing behavior.
The nondiffusing motion is modeled as a sum of oscillations at the
normal mode frequencies (Eq.\ (\ref{uoft})), simulating the trajectory
of an average atom in a complicated single-body potential well that
fluctuates due to the motion of its neighbors.  Self-diffusion is
accounted for in terms of instantaneous transits between wells, which
occur at a temperature-dependent rate $\nu$.  Since this model gives a
simple and straightforward account of the motion itself, it can easily
be used to calculate any single-atom correlation function one wishes;
here we have focussed on the velocity autocorrelation function.  It is
interesting to note that in this model the velocity correlations
persist through a transit, instead of being washed out entirely by the
transit process; we will return to this point below.  The relaxation
of correlations seen by the decay of $\hat{Z}(t)$ arises here from two
distinct processes: Dephasing as a result of the large number of
frequencies in the single-well motion, and transits between wells.
The dephasing effect produces relaxation but not diffusion: It causes
$\hat{Z}(t)$ to decay \cite{chis} but its integral remains zero.  On
the other hand, transits certainly contribute to relaxation (see
\cite{oldvacf}, where they provide the only relaxation mechanism), but
in addition they raise the first minimum of $\hat{Z}(t)$
substantially, increasing its integral and providing a nonzero $D$.

The comparison of this model to MD results is generally quite
positive, particularly for a one-parameter model; the two calculations
of $\hat{Z}(t)$ agree strongly over 1000 timesteps.  In addition, the
match is encouraging over a very large range of temperatures, from
essentially 0 K to $3 T_{m}$.  The most noticeable discrepancies are
the backward drift in the location of the first minimum of
$\hat{Z}(t)$, which is present in MD but not the model, and the
tendency of the model to exaggerate certain characteristics of the MD
results (the maxima at intermediate times and the high-$T$ positive
tail).  This latter effect is responsible for the model's overestimate
of the diffusion coefficient, though we hasten to add that the model
$\hat{D}$ is still in satisfactory agreement with MD results,
especially for the liquid at $T \geq T_{m}$ (see Fig.\ \ref{DvsT}).

As in \cite{chis}, it is useful to compare this model and the
accompanying results to the work others have done using the formalism
of Instantaneous Normal Modes (INM) and similar methods.  Previously, we
discussed the advantages of our general approach and the superior
quality of its results when applied to nondiffusing states; here we
will consider matters relating explicitly to mechanisms of diffusion.
As is noted explicitly by Vallauri and Bermejo \cite{VB}, the account
of INM by Stratt (see, for example, \cite{Stratt}) does not consider
\mbox{diffusion} at all; their $Z(t)$ is essentially a sum over
cosines, and as such it integrates to zero.  This is understandable,
because as Stratt et al.\ repeatedly emphasize, their approximation is
valid only for very short times, so they are not attempting to model
effects with longer timescales.  They compare their \mbox{INM} results with
MD calculations of $\hat{Z}(t)$ for states of an LJ system ranging
from a moderately supercooled liquid to well above the melting
temperature, and our fits are of roughly the same quality or better in
all cases.  Authors who do attempt to model diffusion usually follow
the path suggested initially by Zwanzig \cite{Zwan}, who thought of
the liquid's phase space as divided into \mbox{``cells''} in which each atom
spends its time before finding a saddle point in the potential and
jumping from one cell to the next.  He imagined as a first
approximation that the jumps destroy all correlations between the
cells; since atoms are jumping all the time, he suggested that the net
result was to multiply the nondiffusing form of $Z(t)$ by a damping
factor $\exp(-t/\tau)$ for some timescale $\tau$ representing the
lifetime of a stay in a single cell.  Notice that Zwanzig provided no
dynamical model of the jumping process itself.  This suggestion has
been developed and transformed extensively by Madan, Keyes, and Seeley
\cite{MKS}, who take a general Zwanzig-like functional form for $Z(t)$
and use a combination of heuristic arguments and constraints on its
moments to specify its dependence on a ``hopping rate'' $\omega_{v}$,
the analog of $\tau^{-1}$ for Zwanzig, which they then extract from
the unstable lobe of the INM spectrum.  Although we cannot be entirely
sure, as indicated in \cite{chis}, we think it most likely that their
simulations of LJ Ar at 80 K, 120 K, and 150 K consist of states
comparable to our Figs.\ \ref{216.3Kvsmodel} and \ref{309.7Kvsmodel},
and again we would argue that our fits are somewhat better.  Finally,
Cao and Voth \cite{CV} also approach diffusion by means of a damping
factor, and they consider factors of two different types, each of
which contains parameters that can be determined from other calculated
or experimental quantities.  Their matches with MD are actually quite
good, but again ours are of at least comparable quality.

Having claimed that our matches to MD simulations are as good as or
better than all of the others we have surveyed, let us emphasize a
fundamental difference between our approaches to diffusion: We provide
an account of the {\em process} of transiting from well to well, so we
have a model of the actual {\em motion} of a mean atom in space that
is valid to arbitrary times, and given this model we can {\em
calculate} the effect of transits on correlations.  All other
approaches we know of begin with an account of the motion valid only
for very short times, calculate $Z(t)$ from this motion, and then try
to model the effects of diffusion on $Z(t)$ directly, using parameters
that are thought to be characteristic of jumps between wells.  In the
diffusing regime, no one else we have seen actually calculates
$\langle \mbox{\boldmath $v$}(t) \cdot \mbox{\boldmath $v$}(0)
\rangle$ to find $Z(t)$, as we do.  In the process, we find that some
of the assumptions made by others, in particular Zwanzig's hypothesis
that jumps between wells simply erase correlations, are not true.
This approach is already yielding insights into the actual motion of
an atom undergoing a transit.

Finally, we would like to compare the present mean atom trajectory
model with an earlier independent atom model by Wallace
\cite{oldvacf}.  In developing the independent atom model, two
arguments were made: (a) the high rate of transits in the liquid state
shows the need to abandon the normal mode description of motion, and
instead picture the motion of a single atom among a set of fluctuating
wells, and (b) the leading approximation to a fluctuating well is its
smooth time average well.  Accordingly, the independent atom
oscillates with frequency $\omega$ in a smooth isotropic well, and it
transits with probability $\mu$ to an adjacent identical well at each
turning point \cite{oldvacf}.  What we have now learned by considering
low temperatures, and especially the nondiffusing states of
\cite{chis}, is that the MD system exhibits a velocity decorrelation
process which results from the presence of many frequencies in the
single-particle motion, making assumption (b) less reasonable.  In the
current model these many frequencies are retained, and they are
intended to represent the strong fluctuations in each single-particle
well; in this way the current model makes an important improvement
over the independent atom model, where the well fluctuations were
averaged out.  Beyond this difference, the two models contain similar
but not identical treatments of transits, whose rate increases with
increasing temperature, and which produce self-diffusion.  The less
detailed but simpler independent atom model has proven useful in a
description of the glass transition \cite{glass}, and the
corresponding transit parameter has been used to relate shear viscosity
and self-diffusion in liquid metals \cite{MT}.

The next logical step in this work, given the results so far, is to
extend the mean atom trajectory model and apply it to calculations of
more complicated correlation functions of the liquid's motion.

\end{document}